\documentclass[twocolumn,amsmath,amssymb,superscriptaddress]{revtex4}
\usepackage{soul}
\usepackage{color}
\usepackage[pdftex]{graphicx}
\usepackage{amssymb,amsfonts,amsmath}
\usepackage{soul}
\usepackage{color}

\usepackage{hyperref}
\usepackage{mathtools}
\usepackage{lineno}
\begin{document}

\title{{Long Wavelength Fluctuations and the Glass Transition in 2D and 3D}}
\author{Skanda Vivek} 
\affiliation{Department of Physics, Emory University, 
Atlanta, Georgia 30322, USA}
\author{Colm P. Kelleher} 
\affiliation{Department of Physics and Center for Soft Matter
Research, New York University, New York, New York 10003, USA}
\author{Paul M. Chaikin} 
\affiliation{Department of Physics and Center for Soft Matter
Research, New York University, New York, New York 10003, USA}
\author{Eric R. Weeks}
\affiliation{Department of Physics, Emory University, 
Atlanta, Georgia 30322, USA}
\email{erweeks@emory.edu}

\date{\today}
\begin{abstract}
Phase transitions significantly differ between two-dimensional
and three-dimensional systems, but the influence of
dimensionality on the glass transition is unresolved.  We use
microscopy to study colloidal systems as they approach their
glass transitions at high concentrations, and find differences
between 2D and 3D.  We find that in 2D particles
can undergo large displacements without changing their position
relative to their neighbors, in contrast with 3D.  This is
related to Mermin-Wagner long-wavelength fluctuations that influence phase transitions in 2D.  However, when measuring particle motion only relative to their neighbors, 2D and 3D have similar behavior as the glass transition is approached, showing that the long wavelength fluctuations do not cause a fundamental distinction between 2D and 3D glass transitions.
\end{abstract}
\maketitle

\section*{Introduction}

If a liquid can be cooled rapidly to avoid
crystallization, it can form into a glass:  an amorphous solid.  The
underlying cause of the glass transition is far from clear, although
there are a variety of theories \cite{biroli13,ediger12,cavagna09}.
One recent method of understanding the glass
transition has been to simulate the glass transition in
a variety of dimensions (including 4 dimensions or higher)
\cite{Flennerncomm2015,sengupta12,vanmeel09a,charbonneau10,charbonneau11}.
Indeed, the glass transition is often thought to be similar
in 2D and 3D~\cite{DoliwaPRE2000,hunter12rpp} and in simple
simulation cases such as hard particles, one might expect
that dimensionality plays no role.  As a counterargument,
two-dimensional and three-dimensional fluid mechanics are
qualitatively quite different \cite{tritton88}.  Likewise,
melting is also known to be qualitatively different in 2D and
3D~\cite{BernardPRL2011,StrandburgRevmodphys1988, MaretPRL2000,
Gasser2010melting}.

Recent
simulations give evidence that the glass transition is also quite
different in 2D and 3D \cite{Flennerncomm2015,sengupta12}.
In particular, Flenner and Szamel~\cite{Flennerncomm2015} simulated several different
glass-forming systems in 2D and 3D, and found that the dynamics
of these systems were fundamentally different in 2D and 3D.
They examined translational particle motion
(motion relative to a particle's initial position) and
bond-orientational motion (topological changes of neighboring
particles).
They found that in 2D these two types of motion became decoupled
near the glass transition.  In these cases, particles could move
appreciable distances but did so with their neighbors, so that
their local structure changed slowly.  In 3D, this was not the
case; translational and bond-orientational motions were coupled.
They additionally observed that the transient localization
of particles well known in 3D was absent in the 2D data.
To quote Flenner and Szamel, ``these results strongly suggest
that the glass transition in two dimensions is different than in
three dimensions.''

In this work, we use colloidal experiments to test dimension
dependent dynamics approaching the glass transition.  Colloidal
samples at high concentration have been established as model glass
formers \cite{Pusey86,kegel00,WeeksScience2000,hunter12rpp,MaretRevSc2009}.
We perform microscopy experiments with two 2D bidisperse
systems, one with with quasi-hard interactions, and the other with
long range dipolar interactions.  3D data are obtained from previous
experiments by Narumi \textit{et al.}~\cite{narumisoftmatter2011}
which studied a bidisperse mixture of hard particles.  Our results
are in qualitative agreement with the simulations of Flenner
and Szamel. 

We believe our observations are due to the
Peierls instability \cite{peierls34,landau37}, also called
Mermin-Wagner fluctuations \cite{mermin66,mermin68}.  As Peierls
originally argued, there exist long-range thermal fluctuations in
positional ordering in one-dimensional and two-dimensional solids.
Klix {\it et al.} and 
Illing {\it et al.} recently noted that these arguments should
apply to disordered systems as well \cite{klix15,Illing2016}.  One can
measure particle motion relative to the neighbors of that particle
to remove the influence of these long wavelength fluctuations
\cite{MaretEPL2009}.  Using this method we observe that the
translational and structural relaxations are similar between 2D
and 3D, demonstrating that the underlying glass transitions are
unaffected by the Mermin-Wagner fluctuations.

\section*{Results}

We analyze three different types of colloidal samples, all using 
bidisperse mixtures to avoid crystallization. The first sample type
is a quasi-2D sample with hard particles (short range, purely
repulsive interactions) which we term `2DH.'  The 2DH sample is
made by allowing silica particles to sediment to a monolayer on
a cover slip~\cite{RoyallJPCM2015}.  Our 2DH system is analogous
to a 2D system of hard disks of the sort studied
with simulations~\cite{DoliwaPRE2000,DonevPRL2006}.  The control
parameter is the area fraction $\phi$, with glassy samples found for
$\phi \geq 0.79$.  The second sample type is also
quasi-2D but with softer particles, which we term `2DS.'  The 2DS
system is composed of bidisperse PMMA particles dispersed in oil,
at an oil-aqueous interface~\cite{Kelleher2015}.  The interactions
in this system are dipolar in the far-field limit, and the control
parameter is the dimensionless interaction parameter $\Gamma_{2DS}$,
related to the area fraction.  $\Gamma_{2DS}$ is
defined in the Methods section, with glassy behavior found for 
$\Gamma \geq 530$.  For the third sample type,
`3D,' we use previously published 3D data on a bidisperse sample of
hard-sphere-like colloids \cite{narumisoftmatter2011}.  For these
data, the control parameter is the volume fraction
$\phi$ with glasses found for $\phi \geq 0.58$
\cite{narumisoftmatter2011}.  Details of the sample preparation and
data acquisition for these three sample types are in the Methods
section.  For each sample type the glass transition
is defined as the parameter ($\Gamma$ or $\phi$) above which the
sample mean square displacement (MSD) does not equilibrate in experimental time scales, $\sim 10$
hours for the 2D samples and $\sim 3$ hours for the 3D samples.

\begin{figure}%
\includegraphics[width=\columnwidth]{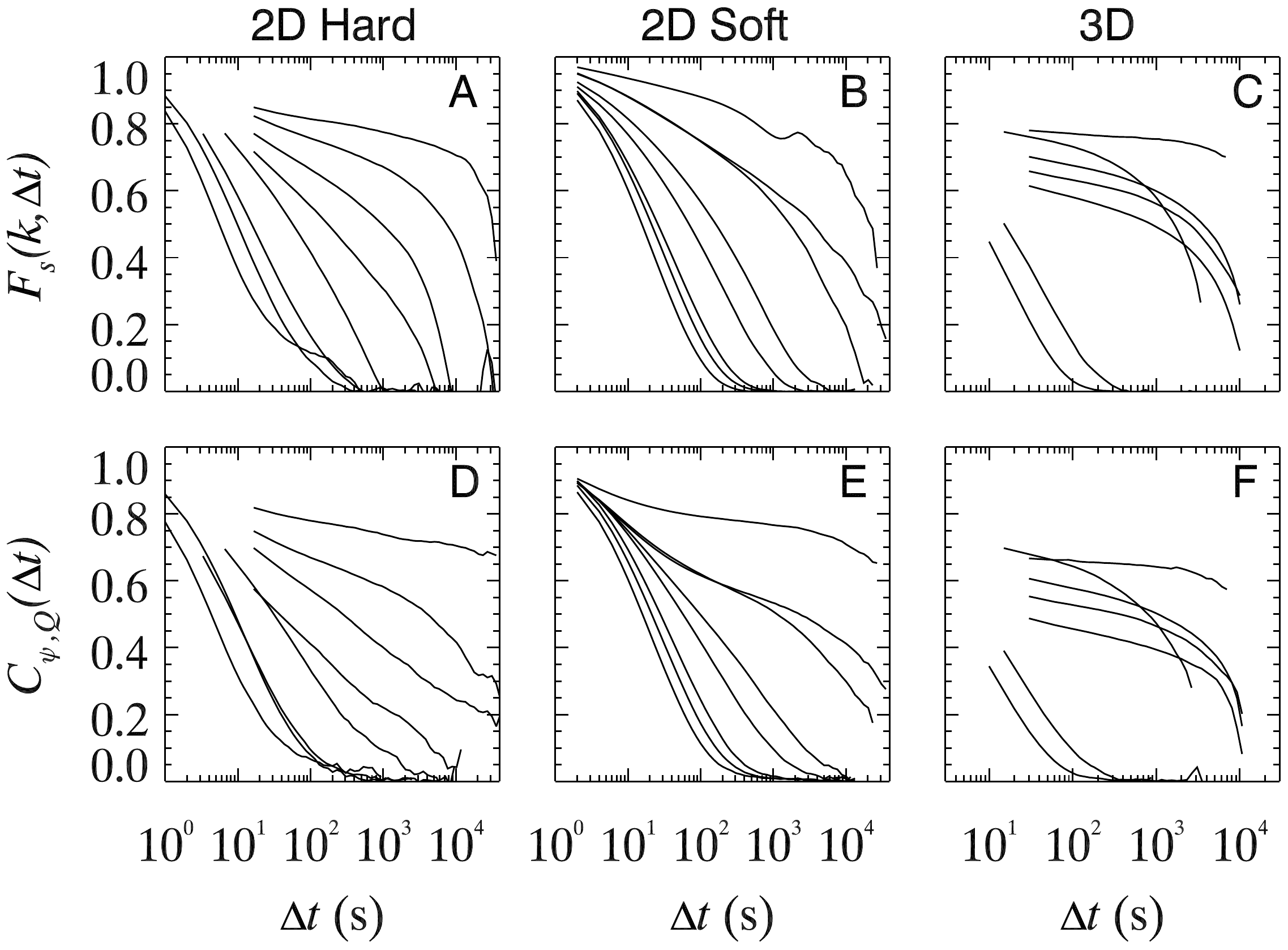}
\caption{Structural relaxation in two and three dimensions.
($A$-$C$)
Self-intermediate scattering functions characterizing
translational motion, using the wave vector $k$ corresponding to
the peak of the structure factor (see Methods and Materials).
($D$-$F$)
Bond-orientational correlation functions.
The columns correspond to 2DH, 2DS, and 3D experiments.  The
parameters for the experiments are:
$\phi_{2DH}=0.55, 0.65, 0.70, 0.74, 0.75, 0.76, 0.78,$ and 0.78;
2DS ($\Gamma_{2DS}=60,100,100,140,180,310,300,$ and  460); 
3D $\phi_{3D}=0.40, 0.42, 0.52, 0.53, 0.54, 0.54,$
and 0.57.  These parameters increase from left to right in each
panel; or equivalently, from bottom to
top.}%
\label{fig:isf_bond}%
\end{figure} 

Flenner and Szamel found that in 2D particles move
large distances without significantly changing
local structure~\cite{Flennerncomm2015}.
They noted that time scales for translational motion
and time scales for changes in local structure were coupled in 3D,
but not in 2D.  The standard way to define
these time scales is through autocorrelation functions.  Following
ref.~\cite{Flennerncomm2015}, 
we compute the self-intermediate
scattering function $F_S(k,\Delta t)$ to characterize translational
motion, and a bond-orientational correlation function
$C(\Delta t)$ to characterize changes in local structural
configuration (see Methods for details).  These are plotted in
Fig.~\ref{fig:isf_bond}$A-C$ and \ref{fig:isf_bond}$D-F$
respectively.  At short time scales, particles have barely moved,
and so both of these correlation functions are close to 1.  At
longer time scales these functions decay, taking longer time
scales to do so at larger concentrations.
The traditional relaxation time scale $\tau_\alpha$ is defined
from $F_S(\tau_\alpha) = 1/e = 0.37$. 
For the bond-orientational correlation functions,
we quantify local arrangements of particles through $\psi_6$ in 2D
and $Q_6$ in 3D, both of which are sensitive to hexagonal order
\cite{NelsonPRB1983}.  Decay of the autocorrelation functions
for these quantities (Fig.~\ref{fig:isf_bond}$D-F$) reflects
how particles move relative to one another, thus changing their
local structure, whereas decay of $F_S$ reflects motion relative
to each particle's initial position.

Specifically, Flenner and Szamel found that $F_S(\Delta t)$ and
$C(\Delta t)$ had qualitatively different decay forms in 2D, but
were similar in 3D
\cite{Flennerncomm2015}.
In particular, $F_S(\Delta t)$ decayed significantly faster than
$C(\Delta t)$ for 2D simulations.  
This means that in 2D particles
could move significant distances (of order their interparticle
spacing) but did so in parallel with their neighbors, so that
their positions were changed but not their local structure.

\begin{figure*}[tb]
\includegraphics[scale=0.99]{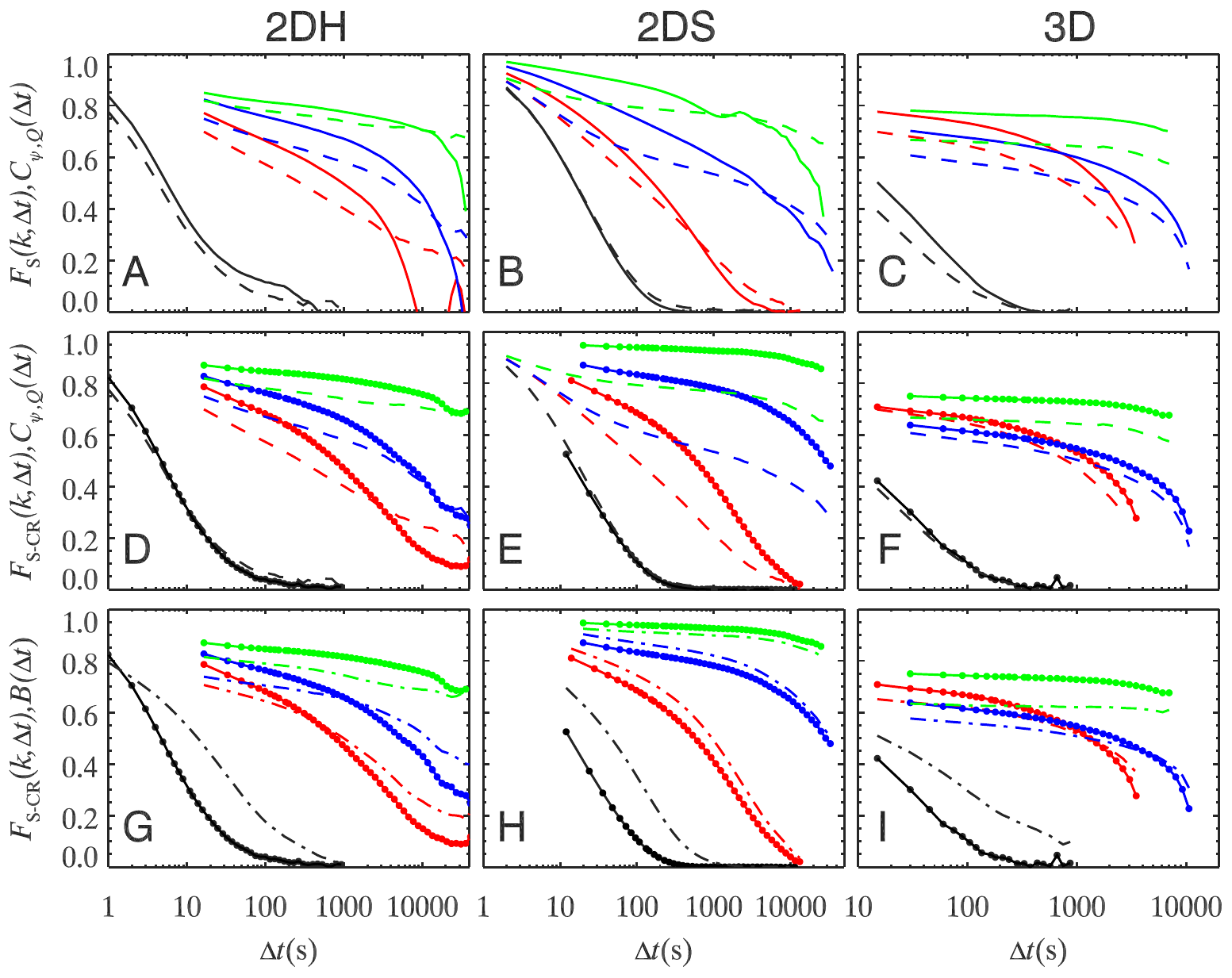}
\caption{
Translational, bond-orientational, and bond-break correlation functions.
($A$-$C$) The solid curves
are $F_S(\Delta t)$ (translational correlations)
and the dashed curves are $C(\Delta t)$ (bond-orientational
correlations) for the 2DH, 2DS, and 3D samples as labeled.
The colors indicate different control parameters.  For 2DH the
colors black, red, blue, and green denote $\phi_{2DH}= 0.55,
0.75, 0.78,$ and 0.78 respectively.  For 2DS the colors black,
red, blue, and green denote $\Gamma_{2DS}= 60, 180, 310,$ and 460
respectively.  For 3D the colors black, red, blue and
green denote $\phi_{3D}= 0.42, 0.52, 0.54,$ and 0.58 respectively.
($D$-$F$) The
solid curves with circles are $F_{S-CR}(\Delta t)$ 
(cage-relative translational correlations).
The dashed curves are $C(\Delta t)$ which are identical to those
shown in ($A$-$C$).
($G$-$I$)
The solid curves with circles are $F_{S-CR}(\Delta t)$ 
(cage-relative translational correlations)
and the dot-dashed curves are $B(\Delta t)$ (bond-break
correlations) for the 2DH, 2DS, and 3D samples.
}%
\label{fig:bond_isfcompare}%
\end{figure*}

To compare translational and bond orientational correlation
functions of our data, we replot some of the data in
Fig.~\ref{fig:bond_isfcompare}$A-C$.  The translational correlation
functions for different parameters are solid curves with different
colors.  The bond-orientational correlation functions are dashed
curves, with same color as corresponding translational correlation
functions. 

The 2D data of Fig.~\ref{fig:bond_isfcompare}$A-B$ exhibit
decoupling, whereas the 3D data of $C$ are coupled.  For the latter
case, coupling means that the 
two functions decrease together, and their
relative positions do not change dramatically as the glass
transition is approached.  Even for the most concentrated case,
for which we do not observe a final decay of either function, it
still appears that the two correlation functions are related and
starting an initial decay around the same time scale.  In contrast,
for both 2D cases (Fig.~\ref{fig:bond_isfcompare}$A,B$), $F_{\rm
S}$ and $C_\psi$ change in relation to one another as the glass
transition is approached.  For 2DH (panel $A$), at the most
liquid-like concentration (black curves), $C$ decays faster than
$F_S$ (dashed curve as compared to the solid curve).  As the glass
transition is approached, initially $C$ decays faster, but then
the decay of $F_S$ overtakes $C$.  A similar trend is seen for 2DS
(panel $B$).  For both 2DH and 2DS, the decoupling is most strongly
seen for the most concentrated samples (green curves), for which
$F_{\rm S}(\Delta t)$ decays on experimental time scales but where 
$C_\psi(\Delta t)$ decays little on the same time scales.

\begin{figure}%
\includegraphics[scale=0.5]{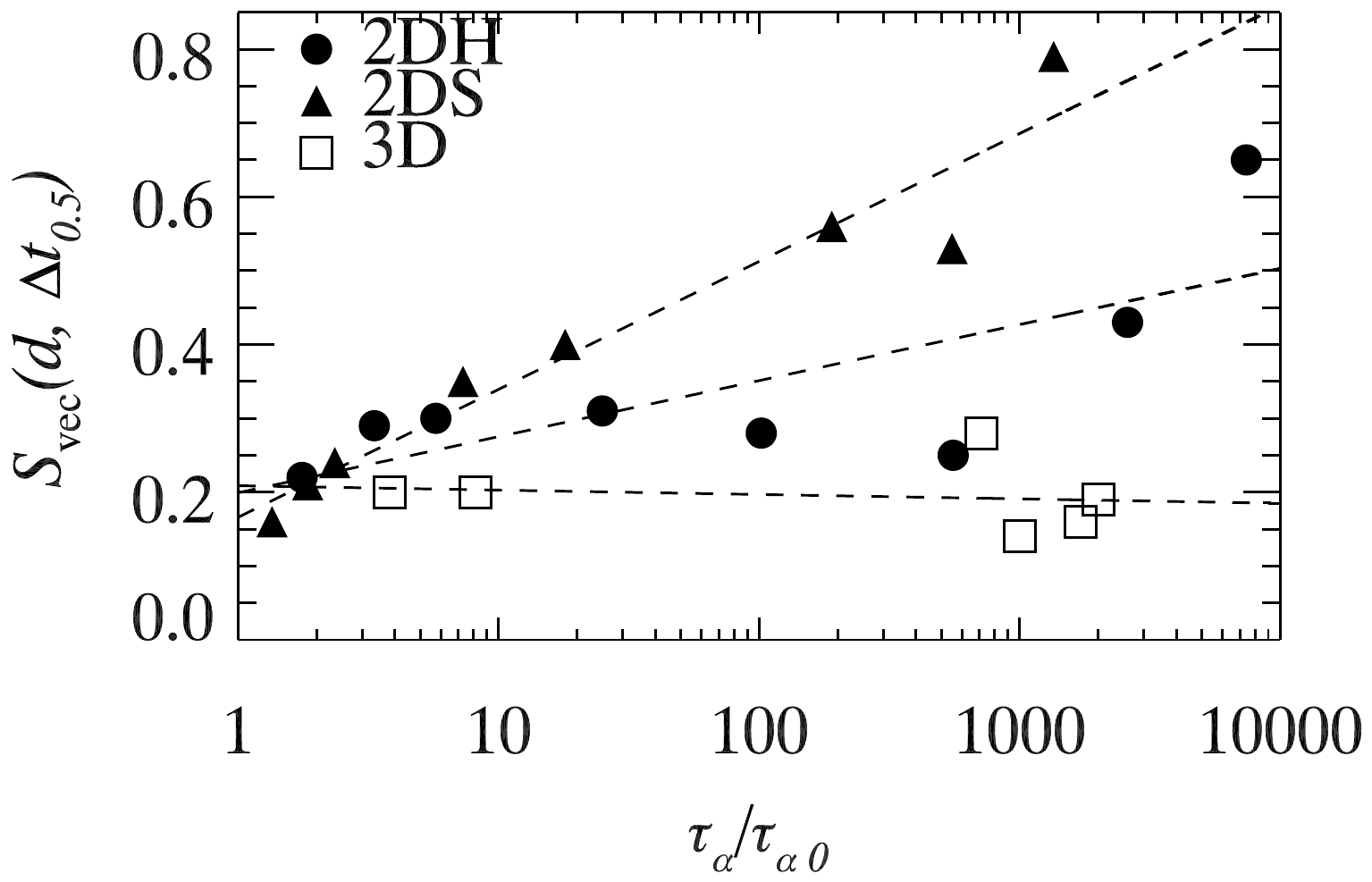}
\caption{
Vector displacement correlations.
The data are for 2DH (filled circles), 2DS (filled triangles),
and 3D (open squares).
The displacements are calculated using a time
scale $\Delta t$ such that $F_S(\Delta t) = 0.5$.
These are measured for all pairs of particles separated by the
nearest neighbor spacing $d$.  $d$ is determined from the
large-large peak position in the pair correlation function
$g(r)$ at the highest concentrations, 
and has values $d=3.38,6.5,$ and 3.10 $\mu$m for 2DH,
2DS, and 3D respectively.
(The location of the $g(r)$ peak depends slightly on $\phi$ for 2DH and 3D
experiments, and more strongly on $\Gamma$ for the 2DS experiments;
for consistency, we keep $d$ fixed to these specific values.)
The lines are least-squares fits to the data.  The data are
plotted as a function of $\tau_\alpha / \tau_{\alpha 0}$ where
$\tau_{\alpha 0}$ is the relaxation time scale for the large
particles in a dilute sample.
2DH (closed circles), 2DS (closed triangles), and
3D (open squares) samples have $\tau_{\alpha 0}=$ 5.4, 20, and
3.8 s respectively.
}%
\label{fig:cor}%
\end{figure}

The slower decay of bond-orientational correlations relative
to translational correlations for our 2D data is in good
qualitative agreement with Flenner and Szamel's observations
\cite{Flennerncomm2015}.  Upon approaching
the glass transition in 2D, particles are constrained to move
with their neighbors such that $C$ decays less than might be
expected on time scales where $F_S$ has decayed significantly.  
In 3D, however, on
approaching the glass transition particles move in
a less correlated fashion.  To quantify the correlated motion
of neighboring particles we compute a two-particle correlation
function \cite{DoliwaPRE2000,WeeksJPCM2007}.  This function
correlates the vector displacements of pairs of nearest
neighbor particles (see Methods).  Fig.~\ref{fig:cor} shows
these correlations: 1 corresponds to complete correlation,
and 0 is completely uncorrelated.  For both 2D samples (solid
symbols) the correlations increase for larger $\tau_{\alpha}$, as
indicated by the fit lines.
This increased correlation reflects particles moving in parallel
directions with their nearest neighbors.  For the 3D data (open
squares in Fig.~\ref{fig:cor}) the correlations are small and do
not grow as the glass transition is approached.  Particle motion
uncorrelated with neighboring particles decorrelates both
positional information and bond-orientational structure.

\begin{figure*}%
\begin{center}
\includegraphics[scale=0.9]{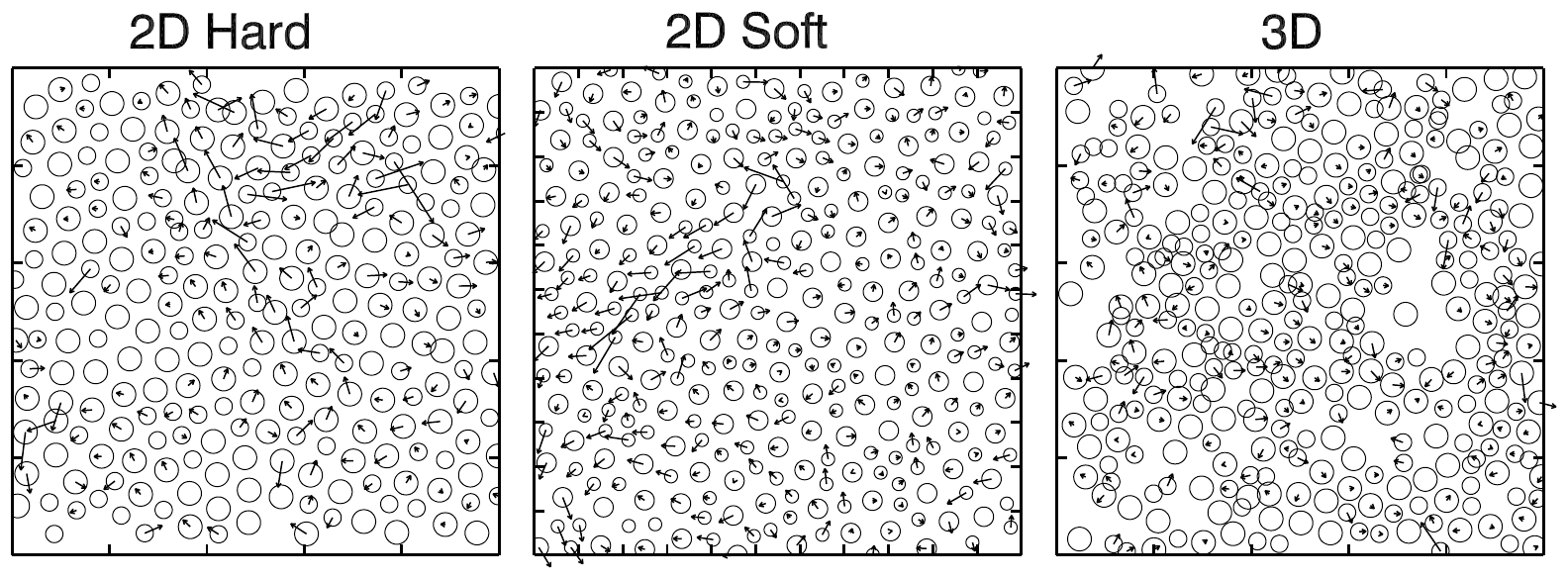}
\caption{
Particle displacements.
These images show displacement vectors of particles using 
a time interval $\Delta t$
chosen such that $F_s(\Delta t) = 0.5$.  For the 3D image
we use an $xy$ cut at fixed $z$.  All scale ticks are at 10
$\mu$m intervals and all displacement vectors are multiplied by two
for easier visualization.
The circles denote particle positions and sizes.
Samples are $\phi_{2DH}=0.78$,
$\Gamma_{2DS}=300$, and $\phi_{3D}=0.54$, from left to
right, with corresponding $\Delta t = 4290, 1720,$ and 3540~s.
$\tau_{\alpha}$ for these samples are 14000, 3800, and 7600~s
respectively.
Circles with no arrows are those with displacements less than 10 \%
of symbol size.
}%
\label{fig:visual}%
\end{center}
\end{figure*}

To qualitatively visualize the differences between dynamics
in 2D and 3D, the top row of Fig.~\ref{fig:visual} shows
displacement vectors for particles in the three samples near
their glass transitions.  For both 2DH and 2DS samples, there
are clusters of particles moving in similar directions as seen
by adjacent displacement arrows pointing in a similar direction.
This clustering is less pronounced in 3D, consistent with the
small correlations between nearest neighbor motions in 3D
(Fig.~\ref{fig:cor}).

As suggested in the Introduction, it is plausible that some of the
significant translational motion in the 2D samples is due to
Mermin-Wagner fluctuations which act at long wavelengths
\cite{Illing2016,Shiba2015}.  To disentangle the potential
influences of long wavelength fluctuations from relative motions,
we subtract collective motions by measuring ``cage
relative'' particle motions \cite{MaretEPL2009}.  The key idea is
to measure displacements relative to the average displacements of
each particle's nearest neighbors, that is, relative to the cage
of neighbors surrounding each particle.  
Previous work has shown that using cage-relative coordinates reveals the dynamical 
signatures of phase transitions for systems of monodisperse colloids~\cite{MaretPRL2000}.
We compute these
cage-relative displacements and then calculate the
self-intermediate scattering function $F_{\rm S-CR}$ using
these new displacements.  These are plotted as solid lines with
circles in Fig.~\ref{fig:bond_isfcompare}$D-F$, with the dashed
lines being the bond-orientational data (which are unchanged as
$C(\Delta t)$ is always calculated relative to neighbors).  In both
2DH and 2DS, $F_{S-CR}(\Delta t)>F_S(\Delta t)$ (the solid
lines in Fig.~\ref{fig:bond_isfcompare}$D,E$ are higher than the
corresponding solid lines in Fig.~\ref{fig:bond_isfcompare}$A,B$).
This is expected given the arguments above, that particles move
with their neighbors, hence subtracting nearest neighbor motions
results in reduction of particle mobility.  For the 3D data
(Fig.~\ref{fig:bond_isfcompare}$F$), the $F_{\rm S-CR}(\Delta t)$
curves still show coupling to $C_{\rm Q}(\Delta t)$ similar to
the original data shown in Fig.~\ref{fig:bond_isfcompare}$C$.

To provide a complementary view, we consider another measure of
structural changes, the cage correlation function (or bond-breaking
function) $B(\Delta t)$.  $B(\Delta t)$ is the fraction of particles
that have the same neighbors at times $t$ and $t+\Delta t$,
averaged over $t$ \cite{rabani97,weeks02}.

These functions are plotted
in Fig.~\ref{fig:bond_isfcompare}$G-I$ as dash-dotted lines,
and are compared to $F_{\rm S-CR}$.  
The black curves are the lowest concentrations, which all have
$B(\Delta t)>F_{S-CR}(\Delta t)$.  This  is because at lower
concentrations, particles can translate a significant amount
without losing neighbors.  However at larger concentrations,
$B(\Delta t)\sim F_{S-CR}(k,\Delta t)$ in all 3 types of samples.
For all three experiments, the two correlation functions look
fairly similar at the three highest concentrations shown in
Fig.~\ref{fig:bond_isfcompare}$G-I$.
In particular, the differences between the 2D and 3D data are
much reduced as compared with the original analysis shown in
panels $A-C$.

In fact, our strongest qualitative evidence for coupling comes from
comparison of the green curves in Fig.~\ref{fig:bond_isfcompare},
which are the samples closest to the glass transition.  In each
case, the correlation functions do not fully decay within our
experimental observation time.  Nonetheless, it is apparent
for the 2D data that the normal self-intermediate scattering
function is beginning a final decay at a time scale for which
the bond-orientational function has not yet begun to decay
(Fig.~\ref{fig:bond_isfcompare}$A,B$).  This is not the case for
the 3D data (panel $C$).  In contrast, all three data sets exhibit
similar behavior at the largest time scales when comparing the
cage-relative $F_{S-CR}(k,\Delta t)$ and $B(\Delta t)$ (panels $G-I$).

\begin{figure}%
\includegraphics[scale=0.5]{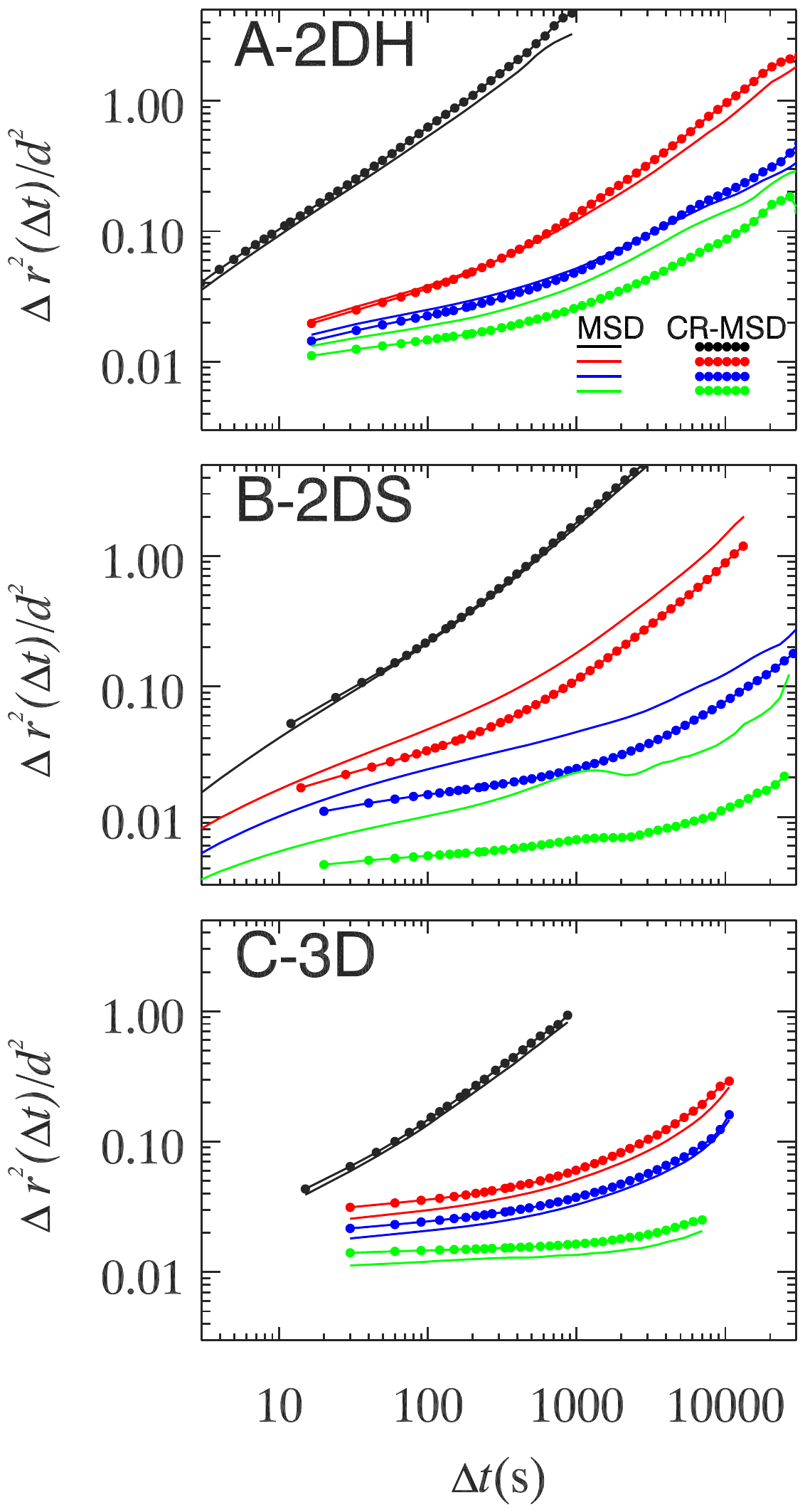}
\caption{
Mean square displacements and cage-relative mean square displacements.
The data $(A-C)$ are for the experiments as indicated.
The solid curves are mean square displacements $<\Delta r^2 >$ 
calculated for all particles, normalized by $d$ as described in
the caption to Fig.~\ref{fig:cor}.
The
solid curves with
circles are cage-relative mean square displacements.
The colors indicate different control parameters, as given 
in Fig.~\ref{fig:bond_isfcompare}. 
For the 3D samples, the $z$ direction
is neglected due to noise and also to facilitate the comparison
with the 2D experiments. 
}%
\label{fig:msd}%
\end{figure}

\begin{figure}%
\includegraphics[scale=0.8]{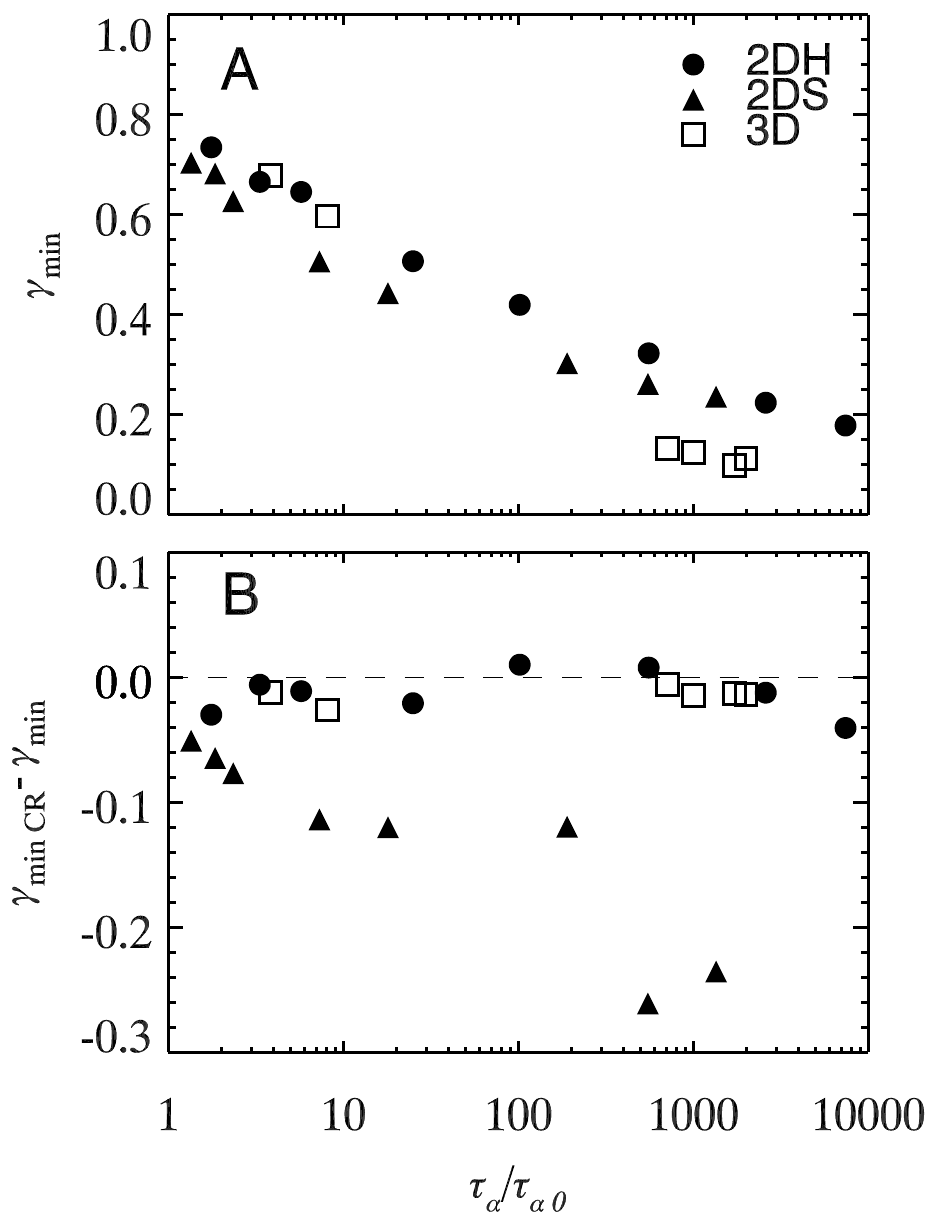}
\caption{
Transient localization parameter.  ($A$)  $\gamma_{\rm min, CR}$ is the
minimum logarithmic slope of the cage-relative mean square
displacements.  
($B$) Difference $\gamma_{\rm min,CR} - \gamma_{\rm min}$ between
the original mean square displacement data and the cage-relative
version.  Negative values indicate the enhancement of measured
transient localization using the cage-relative analysis.
}
\label{fig:gamma}
\end{figure}

We turn now to the question of transient localization, which
Flenner and Szamel found to be present in 3D but not 2D.  
The trajectories of 3D particles showed localized motions
separated by abrupt jumps, while trajectories of 2D particles did
not have these two distinct types of motion
\cite{Flennerncomm2015}.  In their data, this
caused a plateau in the 3D MSD, which was not seen in the 2D MSD.  
The plateau is due to particles being transiently trapped
in cages formed by their neighbors, with the plateau height set
by the cage size \cite{weeks02}.

Motivated by the considerations above, we investigate
the cage-relative mean square displacements (CR-MSD)
\cite{MaretEPL2009}.  In analogy with the cage-relative
scattering function, 
we use the cage-relative displacement $\Delta \vec{r}_{CR}$ to
define the CR-MSD.
Fig.~\ref{fig:msd} shows the original
MSD data (thin lines) and CR-MSD (lines with circles).  For all
experiments as the concentration increases the MSD drops,
reflecting the slowing dynamics on approaching the glass transition.
In some cases, the CR-MSD is larger than the MSD (for example,
all the curves in Fig.~\ref{fig:msd}$C$).  In these situations, the
motion of each particle is less correlated with the motion of its
neighbors, so the cage-relative analysis effectively adds a random
vector to each particle's displacement, thus increasing the MSD
on average.  However, for the 2D samples as they approach the glass
transition, the opposite occurs.  Especially for the green curves in
Fig.~\ref{fig:msd}$A,B$, the data closest to the glass transition,
it is clear that the cage-relative analysis dramatically decreases
the CR-MSD data relative to the original MSD.   While we show data
close to the glass transition, none of our data are from glasses.
There have been
a number of experiments on other 2D colloidal systems such as soft
particles~\cite{yunker2009} and attractive particles~\cite{zhang11}
which observed a slowly rising MSD for glasses.  Our results
suggest that the MSD rise seen in these prior experiments may also
disappear with cage-relative analysis, strengthening the argument
that these prior experiments studied truly glassy samples.

To quantify transient localization, we measure the instantaneous
logarithmic MSD slope $\gamma$ from $\langle \Delta r^2\rangle
\sim \Delta t^{\gamma(\Delta t)}$.  $\gamma = 1$ corresponds to
normal diffusion.  We quantify the amount of localization by
the minimum value of this slope, $\gamma_{\rm min}$; this is
the logarithmic slope at the inflection point of the MSD or
CR-MSD.  Fig.~\ref{fig:gamma}$A$ shows the CR data for the 2D
samples (filled symbols) and 3D (open squares) as a function of
$\tau_{\alpha}$.  While the 3D data reach lower values,
the overall trend is similar between 2D and 3D:  the closer to the
glass transition, the more pronounced transient localization is.
Note that in the work of Flenner and Szamel, they tested both
Newtonian dynamics and Brownian dynamics; the latter is more
appropriate for colloids.  With Brownian dynamics in 2D, they found
slightly more pronounced MSD plateaus.  It is possible that the
presence of Brownian dynamics in our experiments also contributes
to our observed similarities in transient localization between 2D
and 3D.

Fig.~\ref{fig:gamma}$B$ shows the slight enhancement of transient
localization caused by the cage-relative analysis.  We plot the
change in $\gamma_{\rm min}$ upon using the cage-relative analysis,
and it is generally negative.  The largest changes are seen in
the 2DS data (solid triangles), which is sensible as these are
the data with the strongest correlations with their neighbors.

\section*{Discussion}

Our experiments show apparent differences in dynamics approaching
the 2D and 3D colloidal glass transition, in agreement with the
simulation results of Flenner and Szamel \cite{Flennerncomm2015}.

In 2D, we observe that particles move in parallel with their
neighbors, such that their local structure changes less than
if the motions were uncorrelated. While it is clear from prior
work that in 3D particle motions have some correlation with their
neighbors \cite{WeeksJPCM2007}, in our data the correlations are
more significant for the 2D samples.  
These are likely related to Mermin-Wagner fluctuations / the Peierls
instability in 2D
\cite{peierls34,landau37,mermin66,mermin68,klix15,Illing2016}.

Our 2D samples are, of course, quasi-2D.  Both are influenced
by nearby large 3D regions of fluid.  The 2DH sample also has
hydrodynamic interactions between particles and the nearby bottom
of the sample chamber.  We find that 2DS samples are more affected
by long-wavelength fluctuations than 2DH, which could be due to
the difference in interactions \cite{mermin68,frohlich81}. It is certainly
plausible that softer interactions allow for more fluctuations in
the nearest-neighbor distance, whereas for dense samples with
hard interactions, fluctuations are by necessity smaller (as
particles cannot move too close together before they repel)
\cite{Illing2016}.
Recent simulation work has shown
differences in correlation lengths for disks with soft and hard
interaction potentials during 2D melting~\cite{Kapfer2015}.
Nonetheless, the agreement between the two 2D data sets is
striking, especially given the different particle interaction
potentials. Namely as distinct from the 3D samples, both 2D samples
show large Mermin-Wagner fluctuations.

Another important experimental factor is
the system size: approximately $10^5 - 10^6$ for both 2D systems
and $10^9$ for the 3D system.  It is likely that for even larger
2D systems, the Mermin-Wagner fluctuations would be more pronounced
\cite{Flennerncomm2015,Shiba2012,Shiba2015}.

Klix, Maret, and Keim \cite{klix15} recently argued that Mermin-Wagner
fluctuations should be present in glassy systems.  Probably the
most interesting aspect of our study is the suggestion that indeed
2D Mermin-Wagner fluctuations are present in our amorphous samples.
Mermin-Wagner fluctuations conventionally result from elasticity
associated with the development of an order parameter. The origin
of elasticity in glassy systems is less well understood.  While we
have not proven that our observed long-wavelength fluctuations are
indeed Mermin-Wagner fluctuations, one could vary the system size
in future investigations to examine how the difference between
conventional and cage-relative measurements depends on system size.
In conclusion, with our efforts and other recent work, there is a
compelling collection of evidence that
2D and 3D glass transitions are fundamentally the same:  there is 
strong
qualitative agreement between our observations studying three
colloidal systems,
the colloidal experiments and simulations of
Illing {\it et al.} \cite{Illing2016}, and the soft particle
simulations of Shiba {\it et al.} \cite{Shiba2015}.  
The similarities between the conclusions, despite the differences
in methods and dynamics, suggest the results are independent
of the details.
All of these
observations show that the 2D glass transition is 
similar to the 3D glass transition, but with the added influence
of Mermin-Wagner fluctuations in 2D.

\section*{Materials and Methods}

For 2DH experiments, we confine bidisperse non-functionalized
silica particles (diameters $\sigma_S=2.53$ and $\sigma_L=3.38$
$\mu$m, Bangs Laboratories, SS05N) to a monolayer by gravity. 
Prior to taking data, the
sample is quenched by shaking and letting particles sediment on
the coverslip.
The coverslip is made hydrophobic by treatment with Alfa Aesar
Glassclad 18 to prevent particle adhesion.  All particles are
observed to move during the experiment; none adhere to the glass.
We do not add salt.  The
sedimentation lengths for both small ($l_g/\sigma_S=0.019$) and
large particles ($l_g/\sigma_L=0.006$) are small enough to ensure
fast sedimentation and formation of a quasi-2D monolayer; that is,
thermal energy is not enough to overcome the gravitational potential
energy of the particles \cite{hunter12rpp}.  
We verify that in all experiments, only one layer of particles is
present (ensured by keeping the overall particle concentration
below the level that requires a second layer to form).
We use brightfield
microscopy and a CCD camera to record movies of particles diffusing.
This system is analogous to 2D hard disks.  The only caveat is that
the centers of the large and small particles are not at the
same height, so adjacent large and small particles do not contact
each other at their midplane \cite{Thorneywork2014}.


For 2DS, the experimental system is composed of bidisperse
poly-methyl-methacrylate (PMMA) colloids of diameters 1.1 and
2.6 $\mu$m.  The particles are at the interface between oil and a
glycerol/water mixture.  The aqueous phase consists of 10mM NaCl
70 wt.~\% glycerol solution, while the oil phase consists of a
50-30-20 v/v mixture of cyclohexyl bromide, hexane and dodecane.
Interactions between particles are dipolar in the far-field limit. A
dimensionless interaction parameter~\cite{MaretRevSc2009} is used
to characterize the system:
\begin{equation} 
\Gamma_{2DS} =\frac{(\pi
n)^{3/2}}{8 \pi \epsilon k_BT}(\xi p_B+(1-\xi)p_A)^2 \label{eq:2DS}
\end{equation}
where $\epsilon = 4.2\epsilon_0$.  The electric dipole moments are $p_A$
and $p_B= 2300$ and $590$ $e\cdot \mu$m respectively.  $\xi \approx
0.57 - 0.83$ is the number fraction of small particles, and $n$
is the areal density, measured from a Voronoi tessellation.

The 3D sample data were obtained from a previous experiment
by Narumi \textit{et al.}~\cite{narumisoftmatter2011}.  In 3D
experiments, PMMA colloids were stabilized sterically by a thin
layer of poly-12-hydroxy-stearic acid. A binary mixture with
diameters $\sigma_L = 3.10$~$\mu$M and $\sigma_S = 2.36$~$\mu$m
were used.  The number ratio of small particles to large particles
was 1.56.

The imaging regions encompass roughly 400, 1500, and 2000
particles for 2DH, 2DS, and 3D samples respectively at their
highest concentrations.  The total system sizes are much larger,
approximately $10^5 - 10^6$ for both 2D systems and $10^9$
for the 3D system.  We post-processed 2DH and 2DS movies using
particle tracking algorithms~\cite{idlref} to extract particle
positions from individual frames.  The 3D data were previously
tracked using the same algorithm.  Our uncertainty in particle
position is 0.1~$\mu$m for the 2DH experiment, 0.5~$\mu$m for
the 2DS experiment, and 0.2~$\mu$m ($x,y$) and 0.3~$\mu$m ($z$)
for the 3D experiment \cite{narumisoftmatter2011}.

The $\alpha$ relaxation timescales are computed from
self-intermediate scattering functions: $F_S(k,\Delta
t)=\langle\exp(i\vec{k}\cdot \Delta\vec{r})\rangle_t$ where
$\Delta\vec{r}=\vec{r}(t+\Delta t)-\vec{r}(t)$.  The wave vector
$k$ corresponds to the peak of the structure factor 
$S(\vec{k})=\langle N^{-1}|\sum_{i=1}^N \exp(i\vec{k}\cdot\vec{r_i}(t))|^2\rangle$,
where $\vec{r_i}(t)$ denotes particle positions at time $t$ and
the average is over all times.  Corresponding to 2DH, 2DS, and 3D,
$k=2.2,1.0,$ and 2.6 $\mu m^{-1}$, obtained using the average $k$
across all samples of a particular type.

Several other functions we compute require identifying
nearest neighbors, which we do using the Voronoi tessellation
\cite{WeeksScience2000}.

We define cage-relative translational correlation function as:
$F_{S-CR}(k,\Delta
t)=\langle\exp(i\vec{k}\cdot \Delta\vec{r}_{CR})\rangle_t$ where
$\Delta\vec{r}_{CR}=\vec{r}(t+\Delta
t)-\vec{r}(t)-\frac{1}{N}\sum_j[\vec{r_j}(t+\Delta
t)-\vec{r_j}(t)]$,
$j$ denotes nearest neighbors of the particle at initial
time $t$, and the sum is over all neighbors.  The cage-relative
mean square displacement is defined using the same displacements
$\Delta\vec{r}_{CR}$.

To measure bond-orientational correlations in
2D~\cite{Flennerncomm2015}, we define $\Psi_6^n(t)=\sum_m
(N_b^n)^{-1} _m e^{i6\theta_m}$, where $m$ are the nearest
neighbors of particle $n$ and $\theta_m$ is the angle made by
particle $m$ with defined axis.  
From this, the bond-orientational correlation function
can be found as
$C_{\Psi}(\Delta t)=\langle\sum_n [\Psi_6^{n}(t)]^*\Psi_6^n(t+\Delta t)\rangle_t/\langle\sum_n|\Psi_6^n(t)|^2\rangle_t$.

In 3D, we define $Q_{lm}^i(t)=(N_b^i)^{-1}\sum_j
q_{lm}[\theta_{ij}(t),\phi_{ij}(t)]$ where $q_{lm}(\theta,\phi)$ are
spherical harmonics~\cite{NelsonPRB1983,Flennerncomm2015} and the
sum is over neighbors of particle $i$. 
Next we define the correlation function
$Q_l(t_1,t_2)=4\pi/(2l+1)\sum_i\sum_{m=-l}^lQ_{lm}^i(t_2)[Q_{lm}^i(t_1)]^*$.
We calculate $C_Q(\Delta t)=\langle Q_6(t,t+\Delta
t)\rangle_t/\langle Q_6(t,t)\rangle_t$ corresponding to $l=6$,
given that $l=6$ is sensitive to hexagonal order known to be
present even in disordered samples.

The two-particle vector correlations are determined
from a spatial-temporal correlation function defined as
$S_{\textrm{vec}}(R, \Delta t)=\langle\vec{\Delta r_i}\vec{\Delta
r_j} \rangle_{\textrm{pair}}/\langle (\vec{\Delta r^2})\rangle$
\cite{DoliwaPRE2000,WeeksJPCM2007}.  The average is over all
particles with initial separation $R \approx d$, and over
the initial time $t$.  For the initial separation $R$, we use
$R=3.38 \pm 0.2$, $R=6.5 \pm 0.4$, and $R=3.1 \pm 0.2$ $\mu$m
for the 2DH, 2DS, and 3D data.  To determine the displacements
$\Delta \vec{r}$ we use the time scale $\Delta t$ such that
$F_S(\Delta t)=0.5$.  This is chosen to be a shorter time scale
than $\tau_\alpha$, as particle displacements are typically
maximally spatially heterogeneous at a shorter time scale
\cite{WeeksScience2000,weeks02}.

\section*{Acknowledgments}
We thank E.~Flenner, G.~Szamel, R.~Guerra, P.~Keim, H.~Shiba,
and V.~Trappe for useful discussions.  The work of S.V. and
E.R.W. was supported by a grant from the National Science Foundation
(CMMI-1250235).  C.P.K. and P.M.C. were supported by grants from the
National Science Foundation (DMR-1105417), NASA (NNX 13AR67G), and
the MRSEC program of the National Science Foundation (DMR-1420073).


\end{document}